\documentclass[]{aastex63}
\usepackage{graphicx}

\usepackage{amsmath}
\usepackage{natbib}
\usepackage{CJKutf8}
\usepackage{enumitem}
\usepackage{xcolor}

\submitjournal{ApJL}

\shorttitle{Radio Pulses from the M Dwarf}
\shortauthors{Wang et al.}
\graphicspath{{./}{figures/}}

\begin{document}

\title{Rotationally modulated highly circularly polarised radio pulses from the rapidly rotating M dwarf ASKAP J181335-604720}

\correspondingauthor{Shouzhi Wang, Biwei Jiang}
\email{szwang@bao.ac.cn, bjiang@bnu.edu.cn}

\author[0009-0006-8509-8888]{Shouzhi Wang\ 
\begin{CJK}{UTF8}{gbsn}(王守智)\end{CJK}}
\affil{Institute for Frontiers in Astronomy and Astrophysics, Beijing Normal University, Beijing 102206, People's Republic of China}
\affil{School of Physics and Astronomy, Beijing Normal University, Beijing 100875, People's Republic of China}
\affil{National Astronomical Observatories, 
Chinese Academy of Sciences, 
Beijing 100101, People's Republic of China}

\author[0000-0003-3168-2617]{Biwei Jiang\ 
\begin{CJK}{UTF8}{gbsn}(姜碧沩)\end{CJK}}
\affil{Institute for Frontiers in Astronomy and Astrophysics, Beijing Normal University, Beijing 102206, People's Republic of China}
\affil{School of Physics and Astronomy, Beijing Normal University, Beijing 100875, People's Republic of China}

\author[0000-0002-4046-2344]{Qichen Huang\ 
\begin{CJK}{UTF8}{gbsn}(黄启宸)\end{CJK}}
\affil{School of Physics, The University of Sydney, Camperdown
NSW 2050, Commonwealth of Australia}

\author[0009-0005-4091-9980]{Shaoteng Huang\ 
\begin{CJK}{UTF8}{gbsn}(黄少腾)\end{CJK}}
\affil{Institute for Frontiers in Astronomy and Astrophysics, Beijing Normal University, Beijing 102206, People's Republic of China}
\affil{School of Physics and Astronomy, Beijing Normal University, Beijing 100875, People's Republic of China}

\author[0000-0001-6590-8122]{Jundan Nie\ 
\begin{CJK}{UTF8}{gbsn}(聂俊丹)\end{CJK}}
\affil{National Astronomical Observatories, 
Chinese Academy of Sciences, 
Beijing 100101, People's Republic of China}



\begin{abstract}

We report the detection of strong, highly circularly polarised, and rotationally modulated radio pulses from the early--mid M dwarf ASKAP J181335-604720, based on strictly simultaneous radio and optical observations with the Australian Square Kilometre Array Pathfinder (ASKAP) and the Transiting Exoplanet Survey Satellite (TESS). The ASKAP data reveal recurrent broadband radio pulses across 800--1088 MHz, with peak circular polarisation fractions approaching 100\%. A dominant period of $P = 5.607 \pm 0.003$ h is derived from the TESS light curve using a Lomb--Scargle analysis, which we interpret as the stellar rotation period. When phase-folded on this period, the radio emission is confined to narrow phase intervals and recurs at fixed rotational phases, consisting of a dominant pulse and a weaker secondary component. No contemporaneous optical flares are detected at the epochs of the radio pulses in the simultaneous TESS data. Even under conservative assumptions, the inferred brightness temperature exceeds $T_b \gtrsim 1.8 \times 10^{12}$~K, ruling out incoherent emission mechanisms. Combining the observed characteristics, we interpret the emission as coherent electron cyclotron maser (ECM) radiation arising from the stellar magnetosphere, with the observed frequencies implying local magnetic field strengths of at least several hundred Gauss in the radio-emitting region. This work provides a clean, well-constrained, and strictly simultaneous radio--optical case, demonstrating that coordinated radio and optical observations offer a powerful means of distinguishing magnetospheric coherent radio emission from flare-associated coronal activity in M dwarfs.

\end{abstract}

\keywords{stars: activity --- stars: magnetic fields --- pulsars: general --- radio continuum: stars --- radiation mechanisms: nonthermal}

\section{Introduction} \label{sec:intro}

M dwarfs occupy a unique position in stellar astrophysics. They are the most numerous stellar population in the Galaxy \citep{2010AJ....139.2679B} and exhibit some of the most extreme magnetic activity among main-sequence stars \citep{2019A&A...626A..86S, 2021A&ARv..29....1K}. With typical masses of $0.07$--$0.6\,M_\odot$ and effective temperatures of $2500$--$3700$~K \citep{2013A&A...556A..15R, 2024ARA&A..62..593H}, many M dwarfs possess deep convective envelopes, and those at sufficiently low masses become fully convective \citep{2024ARA&A..62..593H}. In combination with rapid rotation, these interiors are thought to enable efficient dynamo action and sustain strong, large-scale magnetic fields \citep[e.g.,][]{2007ApJ...656.1121R, 2017NatAs...1E.184S, 2017ApJ...834...85N, 2021A&ARv..29....1K}. These magnetic fields power activity phenomena spanning radio to X-ray wavelengths, including flares, coronal heating, and large-scale magnetospheric processes \citep{1993ApJ...405L..63G, 2002ARA&A..40..217G}.

Radio emission provides a particularly powerful probe of stellar magnetism, as it directly traces non-thermal particle acceleration and constrains magnetic field strengths and plasma conditions in stellar coronae and magnetospheres \citep{1985ARA&A..23..169D, 2002ARA&A..40..217G}. In M dwarfs, radio emission may arise from incoherent gyrosynchrotron radiation or from coherent processes such as electron cyclotron maser (ECM) emission. The latter can produce extremely high brightness temperatures, strong circular polarisation, and pronounced rotational or quasi-periodic modulation \citep{1982ApJ...259..844M, 1985ARA&A..23..169D, 2006A&ARv..13..229T, 2006ApJ...653..690H, 2007ApJ...663L..25H, 2008ApJ...684..644H, 2019ApJ...871..214V}. Coherent auroral or ECM-like radio emission is observed in a range of magnetised systems, from Solar System planets to magnetic early-type stars and low-mass stars \citep[e.g.,][]{1998JGR...10320159Z,2000A&A...362..281T, 2021NatAs...5.1233C, 2022ApJ...925..125D, 2024ApJ...974..267D}. In the low-mass stellar regime, recent sensitive low-frequency and GHz radio surveys have revealed a growing population of radio-active M dwarfs and ultracool dwarfs \citep{2017ApJ...836L..30L, 2019MNRAS.488..559Z, 2020NatAs...4..577V, 2021NatAs...5.1233C, 2021MNRAS.502.5438P, 2022ApJ...935...99B, 2023A&A...670A.124C, 2026NatAs..10..410T}. In ultracool dwarfs, coherent radio emission is commonly interpreted as auroral ECM emission powered by magnetospheric current systems, analogous to auroral processes in magnetised Solar System planets \citep[e.g.,][]{2007ApJ...663L..25H, 2015Natur.523..568H}. 

For early--mid M dwarfs, however, the physical origin of coherent radio emission can be more ambiguous, because flare-driven coronal activity and magnetospheric processes may both contribute. In the classical flare scenario, radio bursts are generally attributed to non-thermal electrons accelerated during reconnection events, while optical and ultraviolet emission arises from chromospheric and photospheric heating, leading to an expected temporal correlation between the two, albeit sometimes with modest delays \citep{2010ARA&A..48..241B}. Yet an increasing number of low-mass stars exhibit radio behaviour that is difficult to reconcile with purely stochastic flaring, including highly circularly polarised bursts that recur at preferred rotational phases \citep{2006ApJ...653..690H, 2007ApJ...663L..25H, 2019MNRAS.488..559Z,  2023ApJ...951L..43R}. In some systems, such periodic radio emission has also been interpreted in the context of star--planet or star--companion interactions, leaving open the question of whether external drivers are required in addition to intrinsic stellar magnetospheric processes \citep{2007P&SS...55..598Z, 2020NatAs...4..577V, 2021A&A...645A..77P}. The physical origin of coherent radio emission in this regime, including its relationship to classical flaring activity, therefore remains unclear.

In this Letter, we report the radio detection of the early--mid M dwarf ASKAP J181335-604720, based on sub-GHz observations with the Australian Square Kilometre Array Pathfinder (ASKAP) and strictly simultaneous optical photometry from the Transiting Exoplanet Survey Satellite (TESS). These results not only provide key insight into the nature of the radio emission from ASKAP J181335-604720, but also offer a new observational perspective on the magnetic environments and space weather of M dwarfs.

\section{Observations and results}

\subsection{ASKAP Radio Observations}

We use ASKAP observations from the Evolutionary Map of the Universe (EMU) radio continuum survey \citep{2021PASA...38...46N}, which aims to image the southern sky to a typical sensitivity of $\sim20~\mu\mathrm{Jy~beam^{-1}}$ at an angular resolution of $\sim15^{\prime\prime}$. The target source was observed on 2025 June 4 (Epoch 1; SBID 74275) in an observing epoch strictly simultaneous
with the TESS observations, with a total effective on-source integration time of 10 h. The observations cover a frequency range of 800-1088 MHz, with a spectral resolution of 1 MHz and a temporal resolution of 10 s. In addition, as part of the same survey, we obtained an independent second observing epoch on 2025 May 29 (Epoch 2; SBID 74235), which has a comparable frequency coverage, temporal resolution, and on-source integration time. Unless explicitly stated otherwise, the following analysis focuses on Epoch 1, which is strictly simultaneous with the TESS observations.

We also systematically inspected the phase coverage and catalogue detections of other ASKAP epochs at the target position. This search identified two additional epochs with possible detections, SBID 78606 and SBID 36555. However, after reprocessing the data and inspecting the corresponding dynamic spectra, we find that these candidates are not reliable astrophysical detections. Their dynamic spectra show irregular time-frequency structures and contamination patterns, suggesting that they are likely affected by radio frequency interference or imaging artefacts. We therefore do not include them in the main analysis or use them to constrain the rotational emission phase.

The calibrated visibility data were retrieved from the CSIRO ASKAP Science Data Archive (CASDA) in the form of measurement sets and further processed using the \texttt{DSTOOLS} package\footnote{https://github.com/askap-vast/dstools}. Imaging and the corresponding model components were generated separately for Stokes $I$ and Stokes $V$ using \texttt{WSClean} \citep{2014MNRAS.444..606O}. To isolate the time-variable radio emission, a mask was applied at the source position, and the corresponding model visibilities were subtracted from the calibrated data. The residual visibilities were then phase-rotated to the target position and averaged over all baselines to construct the dynamic spectra.

Figure \ref{figa1} shows the time and frequency averaged ASKAP continuum images of ASKAP J181335-604720 in Stokes $I$ and Stokes $V$ for Epoch 1, constructed using the full observing bandwidth and integration time. In Epoch 1, the source is detected with peak flux densities of \(0.70 \pm 0.02\) mJy beam\(^{-1}\) in Stokes \(I\) and \(-0.46 \pm 0.02\) mJy beam\(^{-1}\) in Stokes \(V\). Using the local image rms noise, these measurements correspond to detection significances of \(25\sigma\) and \(24\sigma\), respectively. The source is unresolved at the synthesized beam size of \(17.7^{\prime\prime} \times 15.0^{\prime\prime}\). The corresponding absolute fractional circular polarisation is \(66\% \pm 3\%\). For Epoch 2, the source is also detected in the time and frequency  averaged ASKAP images, with peak flux densities of \(0.45 \pm 0.02\) mJy beam\(^{-1}\) in Stokes \(I\) and \(0.20 \pm 0.02\) mJy beam\(^{-1}\) in Stokes \(V\). Based on the local image rms noise, these measurements correspond to detection significances of \(21\sigma\) and \(7\sigma\), respectively. The corresponding absolute fractional circular polarisation is \(44\% \pm 5\%\).

The middle and lower panels of Figure \ref{fig2} present the radio light curves in Stokes $I$ and Stokes $V$, and the dynamic spectrum in Stokes $V$, respectively. The radio emission consists of recurrent, short-duration radio pulses detected across the full 800-1088 MHz band and appearing synchronously in Stokes I and Stokes V , with peak circular polarisation fractions approaching 100\%, indicative of a highly ordered emission process. The main pulse has a full width at half maximum (FWHM) of about 20 min, much shorter than the recurrence timescale of about 5.6 h, corresponding to a duty cycle of about 6\% and indicating a strongly beamed emission geometry.

\subsection{Identification of the Optical Counterpart}

Based on the ASKAP radio continuum imaging, we determined the peak radio position of ASKAP J181335-604720 to be R.A. = 18:13:35.53 and Decl. = -60:47:20.24 (J2000). The expected astrometric uncertainty of the ASKAP position is conservatively estimated to be $\lesssim2^{\prime\prime}$ \citep{2021PASA...38...46N}. We therefore cross-matched this position against the Gaia DR3 catalogue \citep{2023A&A...674A...1G} using a matching radius of $2^{\prime\prime}$, within which a single Gaia source was identified. The Gaia DR3 source has an astrometric position at epoch J2016.0 of
R.A. = 18:13:35.56 and Decl. = -60:47:18.28 (source ID: 6635080884360470656), corresponding to an angular separation of $\Delta\theta = 1.97^{\prime\prime}$.

Because the ASKAP observations were obtained at a later epoch (2025 June 4), we propagated the Gaia position from J2016.0 to the ASKAP observing epoch using the reported proper motions,
$\mu_{\alpha^\ast} = -22.26~{\rm mas~yr^{-1}}$ and
$\mu_{\delta} = -50.19~{\rm mas~yr^{-1}}$.
After correcting for proper motion, the angular separation between the radio and optical positions is reduced to $\Delta\theta = 1.49^{\prime\prime}$.

Given the ASKAP synthesized beam size of $17.7^{\prime\prime} \times 15.0^{\prime\prime}$, the measured angular offset corresponds to $\sim$10\% of the beam and lies well within the expected ASKAP positional uncertainty. Together with the presence of a single Gaia source within the matching radius and the improved positional agreement after proper-motion correction, we identify this Gaia source as the optical counterpart of ASKAP J181335-604720. This association is further supported by the detection of an X-ray counterpart in the eROSITA all-sky survey. The source is identified as 1eRASS J181335.5-604720 in the DR1 catalogue \citep{2025A&A...704A.344S}, and its X-ray counterpart is independently matched to the same Gaia DR3 source. The identification is additionally corroborated by the consistency observed in subsequent optical and radio periodicity analyses. According to the stellar parameters reported by \citet{2023A&A...674A...1G}, the source has an effective temperature of $T_{\rm eff}=3405$ K, a surface gravity of log $g$ = 4.43, and a distance of 84 pc, consistent with an early--mid M dwarf (M3--M4). In addition, the eROSITA all-sky survey reports a 0.2--2.3 keV X-ray flux of \(F_X = 2.68 \times 10^{-13}~\mathrm{erg~s^{-1}~cm^{-2}}\) for the source \citep{2024A&A...682A..34M}. At the Gaia distance of 84 pc, this corresponds to an X-ray luminosity of \(L_X = 2.27 \times 10^{29}~\mathrm{erg~s^{-1}}\).

\subsection{Simultaneous TESS observations}

Optical photometry strictly simultaneous with the ASKAP radio observations was obtained from the TESS \citep{2015JATIS...1a4003R}. TESS observes the sky in a sequence of 27 day observing sectors. ASKAP J181335-604720 (TIC 365466012) was observed in Sector 93, spanning 2025 June 3 to June 29, which overlaps strictly with our ASKAP radio observations acquired on 2025 June 4.

We retrieved the calibrated TESS light curves from the Mikulski Archive for Space Telescopes (MAST)\footnote{\url{https://archive.stsci.edu/tess/}} and adopted the pre-search data conditioning simple aperture photometry (PDCSAP) product generated by the SPOC pipeline \citep{https://doi.org/10.17909/t9-nmc8-f686}. The PDCSAP light curve has a cadence of 120 s and includes corrections for common instrumental systematics. All data points with non-zero \texttt{QUALITY} flags were excluded from the analysis.

\begin{figure*}[ht!]
\centering
\includegraphics[width=0.99\textwidth]{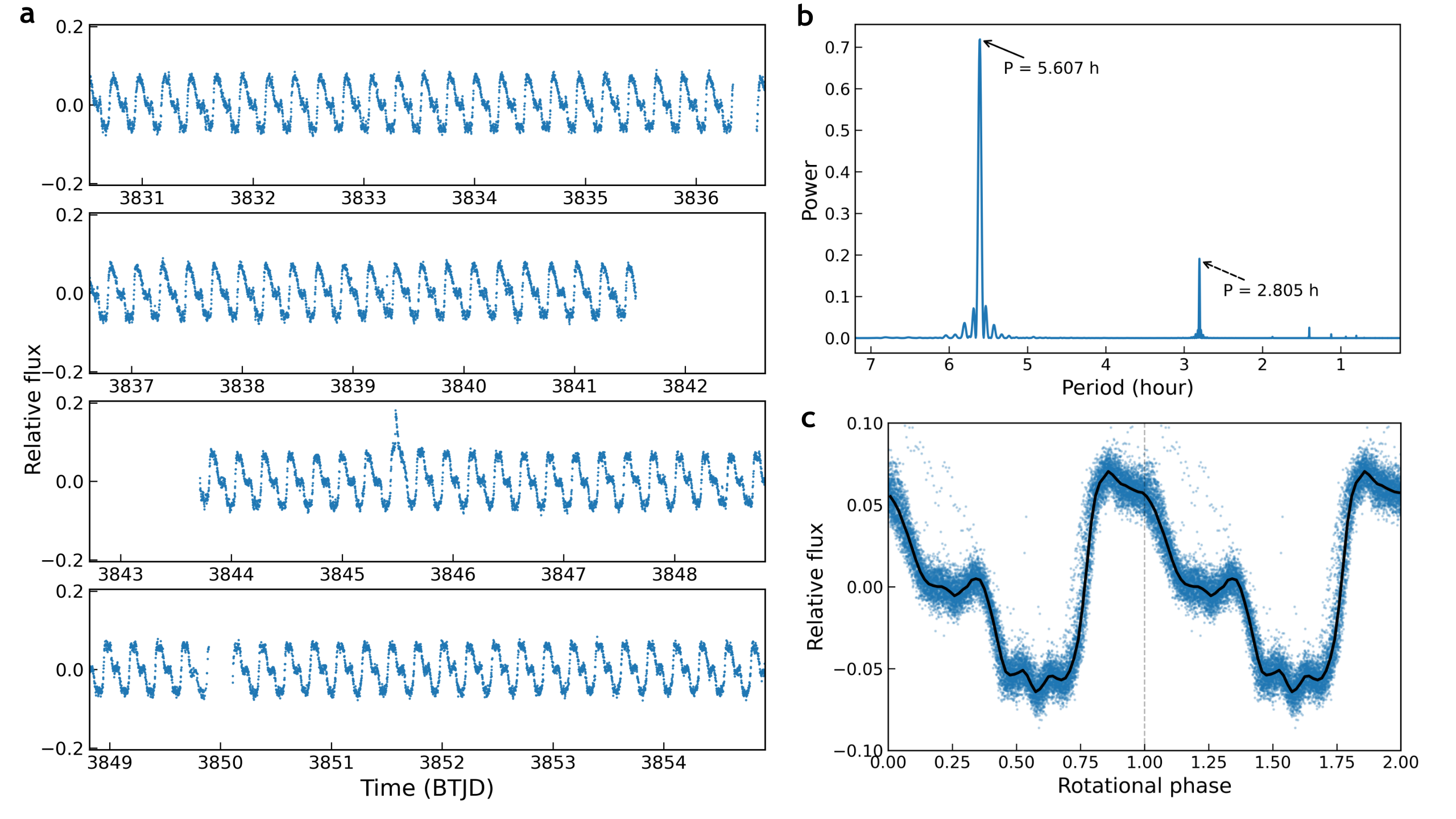}
\caption{TESS photometric analysis of ASKAP J181335-604720 (TIC 365466012).
\textbf{a}, Sector 93 TESS PDCSAP light curve after excluding cadences with non-zero QUALITY flags, shown in four consecutive time intervals for clarity. The flux has been normalized relative to its median value. \textbf{b}, Lomb--Scargle periodogram of the Sector 93 light curve, computed over a total baseline of 24.39 d. A dominant peak is detected at a period of $P = 5.607 \pm 0.003$ h, while a secondary peak at $P = 2.805$ h, close to $P/2$, corresponds to the first harmonic of the rotational modulation. \textbf{c}, TESS light curve phase-folded on the derived rotation period and repeated over phases 0--2 for visual clarity. Phase zero is defined by \(T_0=3831.00474~{\rm BTJD}\) (\(60830.50474~{\rm MJD}\)), corresponding to the start time of the strictly simultaneous ASKAP observation. The black curve shows the median-binned modulation, and the vertical dashed line marks the boundary between the two repeated cycles.}
\label{fig1}
\end{figure*}

The light curve was subsequently normalised by its median flux and converted to fractional units to allow direct comparison with the radio time series (see Figure \ref{fig2}). During the time interval strictly contemporaneous with the ASKAP observations, the optical light curve evolves smoothly following its underlying modulation, with a peak-to-trough variability amplitude of order $\sim$15\% (see Figure \ref{fig2}). In the broader TESS data, however, we identify occasional optical flare-like events, including a noticeable event near BTJD~3845.5 in Sector~93 (see Figure~\ref{fig1}) and several smaller events in other available sectors. This indicates that ASKAP J181335-604720 shows evidence of optical flaring activity, although only a small number of such events are seen in the available TESS data.

To estimate the dominant  timescale of the optical variability, we performed a period analysis of the TESS photometry using the Lomb--Scargle method \citep{1976Ap&SS..39..447L,1982ApJ...263..835S}. The analysis was based on the PDCSAP light curve from TESS Sector~93, after excluding all data points with non-zero \texttt{QUALITY} flags. The dominant period was identified as the peak with the highest power in the periodogram, yielding a period of $P=5.607 \pm 0.003$~h (see Figure \ref{fig1}). The period uncertainty was estimated using bootstrap resampling. We generated 300 bootstrap realizations by resampling the TESS light curve with replacement and recomputed the Lomb--Scargle periodogram for each realization. The quoted uncertainty is half of the 16th--84th percentile range of the resulting distribution of peak periods. We did not combine the other available TESS sectors to derive the adopted period, because we find that the light curve amplitudes and morphologies differ between sectors. These differences may arise from both intrinsic evolution of the spot-modulated signal and sector-dependent TESS systematics or detrending effects. Directly combining widely separated sectors could therefore mix different light curve structures and affect the period analysis. Instead, we performed the Lomb--Scargle analysis separately for each available TESS sector. Since the inferred rotation period (\(P = 5.607\)~h) is much shorter than the duration of an individual TESS sector (\(\sim 27\) days), each sector contains many rotation cycles and can independently constrain the period. As a consistency check, the dominant periods obtained from the individual sectors are consistent with the Sector~93 value, with a maximum difference of only \(0.005\)~h from \(5.607\)~h.

Using the derived period, we phase-folded the TESS photometry to examine the morphology of the optical variability. Rotational phases were computed using
\[
\phi = \left[\frac{t - T_0}{P}\right] \bmod 1 ,
\]
where \(t\) is the TESS observation time in BTJD, \(T_0 = 3831.00474~{\rm BTJD}\), corresponding to \(60830.50474~{\rm MJD}\), is the start time of the strictly simultaneous ASKAP observation and defines phase zero, and \(P=5.607~{\rm h}\) is the stellar rotation period. The modulo operation folds the time series onto a single rotation cycle, so that \(\phi\) ranges from 0 to 1. This choice of phase zero is adopted only to facilitate comparison between the TESS and ASKAP observations and has no specific physical interpretation. The folded light curve exhibits a smooth and highly repeatable quasi-sinusoidal waveform, consistent with rotational modulation caused by surface inhomogeneities such as starspots (see Figure \ref{fig1}), and typical of rapidly rotating M dwarfs with strong large-scale magnetic fields \citep{2014ApJS..211...24M,2020AJ....159...60G}. To highlight the repeatability over multiple rotation cycles, we display the phase-folded light curves over two consecutive cycles (phase 0--2). When the ASKAP radio data are folded on the optical rotation period, the radio emission is confined to a narrow range of rotational phases and recurs at fixed rotational phases (see Figure \ref{fig3}). We then apply the same folding procedure to ASKAP observations obtained in the second independent epoch. Although the overall radio flux density in Epoch 2 is lower, the radio emission remains concentrated at the same rotational phases as observed in Epoch 1 (see Figure \ref{fig3}), indicating that the rotational phase behaviour is reproducible across the two epochs, while the pulse amplitude is variable.

\subsection{Optical flare search and quantitative limits}

To assess whether the coherent radio pulses are associated with impulsive optical flaring, we performed a quantitative flare search on the strictly simultaneous TESS photometry. We used the PDCSAP light curves, and excluded all data points with non-zero quality flags.

To isolate short-duration impulsive events, we removed the long-term rotational variability using a Savitzky--Golay filter \citep{1964AnaCh..36.1627S}, which performs a local polynomial regression within a sliding time window. The filter window length was set to 60 minutes, ensuring that rotational modulation is effectively removed while preserving potential impulsive brightenings on minute timescales. The flare search was conducted on the detrended residual light curve (see Figure \ref{fig:A2}).

The noise properties of the detrended light curve were characterised using a median absolute deviation (MAD) estimator, which was converted to an equivalent Gaussian standard deviation via $\sigma = 1.4826 \times {\rm MAD}$. Potential flare events were identified by searching for positive excursions exceeding a  $3\sigma$ threshold, with the additional requirement that at least two consecutive data points satisfy this criterion, in order to suppress spurious detections caused by single-point outliers.

During the strictly simultaneous interval, no impulsive optical events satisfying these criteria were detected. After removal of the long-term rotational modulation, a single isolated data point exceeding the $3\sigma$ level is present in the residual light curve. However, according to our pre-defined flare detection criterion requiring at least two consecutive points above threshold, this excursion is not classified as an optical flare.


\section{Physical Interpretation of the rotationally modulated Radio Pulses}
\label{sec3}

\begin{figure*}[ht!]
\center

\includegraphics[width=0.9\textwidth]{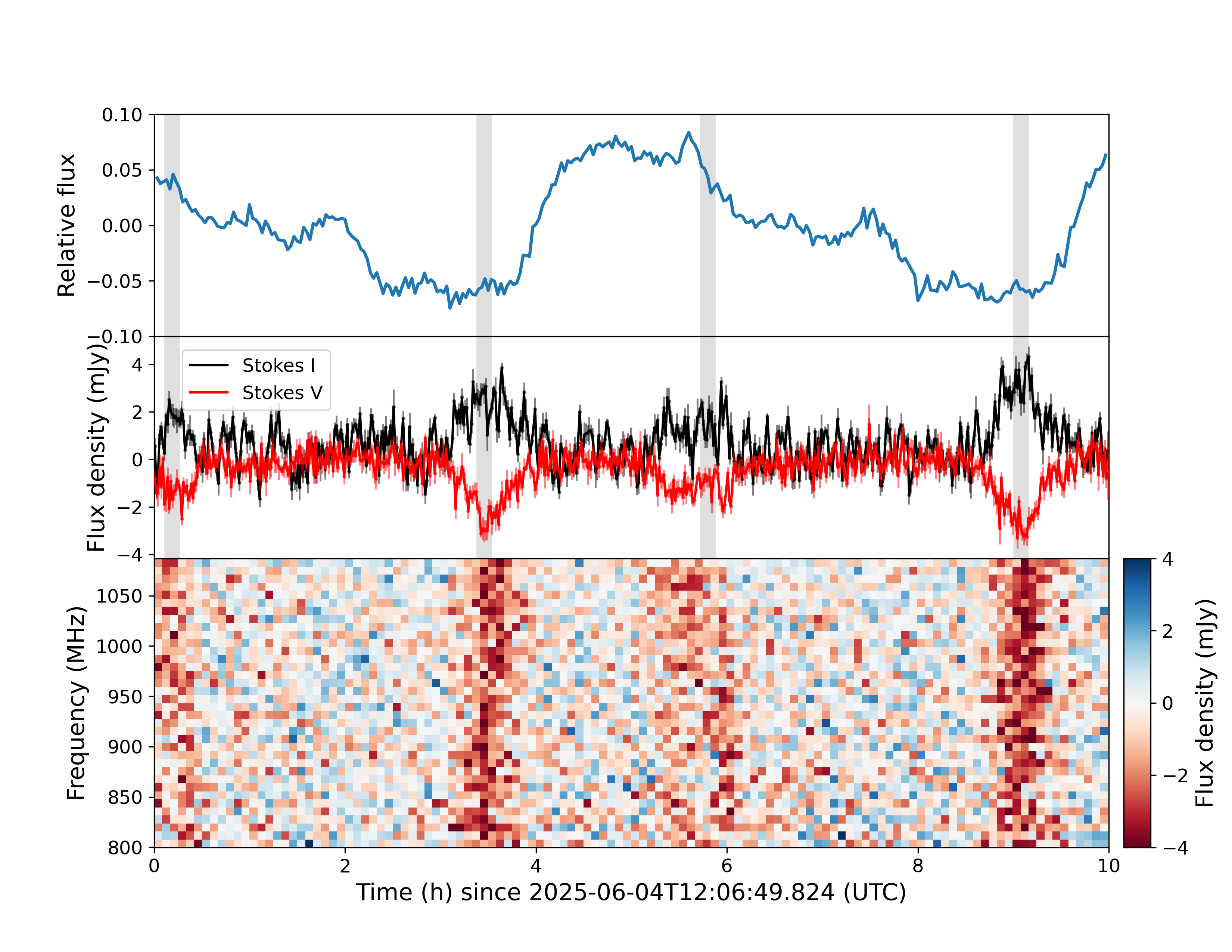}

\center
\caption{Simultaneous ASKAP radio and TESS optical observations of ASKAP J181335-604720. Top panel: The TESS relative optical light curve during the ASKAP observing interval, with the flux normalized to the median value and expressed in fractional units. Middle panel: ASKAP radio light curves in Stokes $I$ (black) and Stokes $V$ (red), binned to 60~s for clarity. Bottom panel: Dynamic spectrum of the ASKAP Stokes $V$ radio emission, binned to 5 min in time and 8 MHz in frequency for clarity. The grey shaded regions indicate the epochs at which radio pulses are detected.}
\label{fig2}
\end{figure*}

\begin{figure*}[ht!]
\centering
\includegraphics[width=0.8\textwidth]{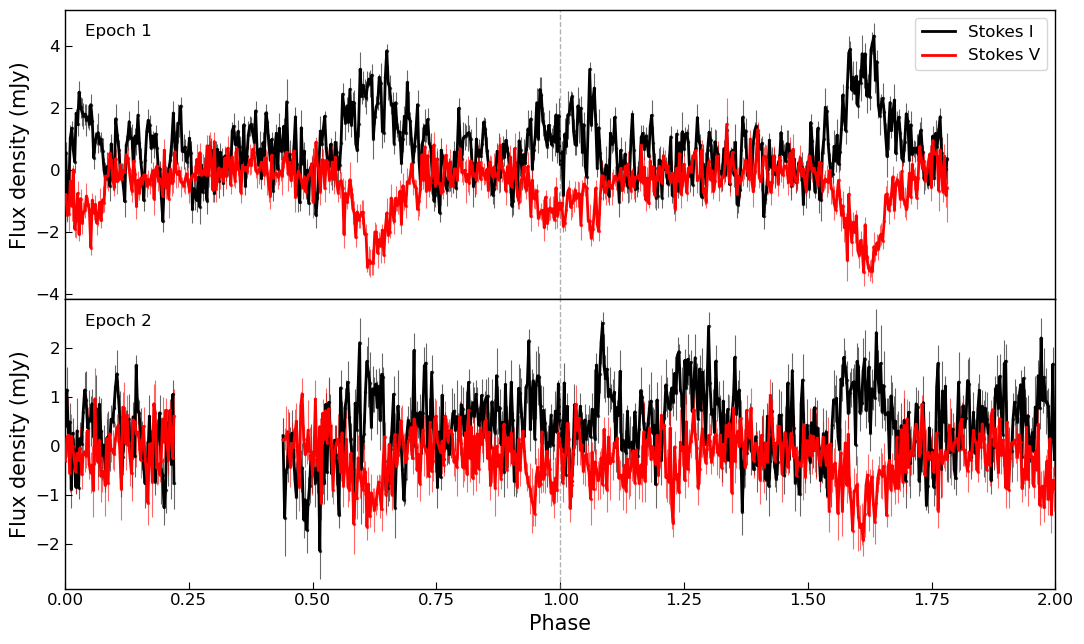}
\caption{Phase-folded ASKAP radio light curves from two observing epochs. The ASKAP Stokes~$I$ (black) and Stokes~$V$ (red) time series are phase-folded using the optical rotation period $P = 5.607$~h derived from the TESS photometry, with the start time of ASKAP Epoch~1 adopted as the reference phase zero-point. The top and bottom panels show Epoch~1 and Epoch~2, respectively.}
\label{fig3}
\end{figure*}

One of the key observational results of this work is that the radio pulses detected by ASKAP exhibit extremely high circular polarisation, with peak fractions approaching 100\%, and recur at stable and well-defined rotational phases (see Figures~\ref{fig3}). At the same time, no evidence for impulsive optical flaring is found at the epochs of the radio pulses (see Figures~\ref{fig2} and \ref{fig:A2}). In the classical stellar flare scenario, radio emission at meter--centimeter wavelengths is generally interpreted as a by-product of non-thermal electrons accelerated during magnetic reconnection, whereas the associated optical emission arises from heating of the chromosphere and photosphere (e.g., \citealt{2003ApJ...597..535H,2010ARA&A..48..241B}). In such cases, the radio and optical emission are usually temporally correlated, although modest delays may occur between them. The absence of optical flares in the strictly simultaneous observations therefore indicates that the radio emission is unlikely to be driven primarily by high-energy flaring events of the kind commonly seen in classical active M dwarfs \citep{1991ApJ...378..725H,2005ApJ...621..398O, 2010ARA&A..48..241B}.

The phase-folded radio light curve shows that the ASKAP pulses do not occur randomly, but instead repeatedly cluster within narrow and well-defined phase intervals (see Figure~\ref{fig3}), comprising a dominant main pulse and a weaker secondary pulse at a different rotational phase. Such behaviour is difficult to reconcile with stochastic flaring activity, because stellar flares do not generally maintain a persistent preference for a single rotational phase. Even in the presence of long-lived active regions, the triggering of individual flares remains intrinsically stochastic, and is therefore unlikely to produce radio pulses that remain confined to the same narrow phase interval over multiple rotation cycles \citep{2014ApJ...797..121H, 2019MNRAS.489..437D, 2019LRSP...16....3T}. Moreover, flares energetic enough to generate the observed radio emission would be expected to leave detectable signatures in the high-cadence, high-precision TESS photometry, which are not observed \citep{2016ApJ...829...23D, 2020AJ....159...60G}.

In addition, we examined a second ASKAP observing epoch (Epoch 2), obtained several days before the strictly simultaneous ASKAP-TESS epoch (Epoch 1). Although Epoch 2 has no strictly simultaneous TESS coverage, it provides an important test of whether the radio emission is recurrent rather than confined to a single observing epoch. As shown in Figure~\ref{fig3}, ASKAP J181335-604720 is again detected in circularly polarised radio emission in Epoch 2. When folded on the same period, the emission is not randomly distributed in rotational phase, but is concentrated within a limited phase range, with phase behaviour similar to that seen in Epoch 1. At the same time, the pulse amplitudes differ between Epoch 1 and Epoch 2, indicating that the emitting region, beaming conditions, or energetic electron population may vary between different observing epochs. Thus, Epoch 2 strengthens the case for rotationally modulated coherent emission, while also showing that the detailed radio properties are not strictly time-invariant.

We also note an apparent phase relationship between the radio pulse phases and the extrema of the optical rotational modulation. The dominant pulse occurs close to the optical flux minimum, whereas the weaker secondary pulse appears closer to the optical flux maximum (see Figures~\ref{fig2} and \ref{fig3}). In the standard interpretation of rotationally modulated optical light curves, such extrema are usually attributed to changes in the projected visibility of surface inhomogeneities, such as cool starspots or magnetically active regions \citep{2005LRSP....2....8B,2009A&ARv..17..251S}. However, this apparent phase relationship should not be interpreted as establishing a reliable correspondence between optical spots and the radio-emitting regions. The optical modulation depends primarily on the distribution of surface brightness inhomogeneities and the stellar inclination, whereas auroral ECM emission is expected to depend on the viewing geometry of the magnetic field, the beaming direction, and the local magnetic-field strength in the emitting region. The apparent phase relationship therefore suggests at most that both the optical and radio variability are modulated by stellar rotation and may be influenced by the same large-scale magnetic topology. The available data do not permit a reliable geometric interpretation in terms of the stellar inclination, magnetic obliquity, or location of the radio-emitting region. 

Nevertheless, the combination of highly circularly polarised radio pulses confined to fixed rotational phases, their repeatability over multiple rotation cycles, and their recurrence with similar phase behaviour in an independent observing epoch strongly favours a coherent radio emission scenario controlled by a stable large-scale magnetic topology, rather than flare emission triggered stochastically by local magnetic reconnection. Similar rotationally modulated radio pulses have been reported in several rapidly rotating low-mass stars and ultracool dwarfs, where they are generally interpreted as coherent emission beamed and controlled by large-scale magnetic structures (e.g., \citealt{2007ApJ...663L..25H,2009ApJ...695..310B,2015ApJ...815...64W,2019ApJ...871..214V}).

\section{Radio emission mechanism and magnetic-field constraints}

The observational results established above place strong constraints on the physical origin of the radio emission from ASKAP J181335-604720 and on the magnetic field strength in the emitting region. In particular, the combination of extremely high circular polarisation, recurrence at well-defined rotational phases and the absence of contemporaneous optical flaring in the strictly simultaneous TESS observations strongly points to a coherent emission mechanism. Following \citet{1985ARA&A..23..169D}, the brightness temperature can be written as
\begin{equation}
T_b = \frac{1}{2k_{\rm B}} \left(S_\nu \times 10^{-26}\right)
\left(\frac{c}{\nu}\right)^2
\left(\frac{d}{l}\right)^2 ,
\label{eq:tb}
\end{equation}
where $k_{\rm B}$ is the Boltzmann constant, $S_\nu$ is the flux density in Jy, $c$ is the speed of light, $\nu$ is the observing frequency, $d$ is the source distance, and $l$ characterizes the linear size of the emitting region.

To obtain a conservative estimate, we assume that the emitting region spans the full stellar disk, i.e., $l = 2R_\star$, and adopt a stellar radius of $R_\star = 0.53\,R_\odot$ from \citet{2022A&A...657A...7K}. Under this assumption, using the observed peak Stokes~$I$ flux density of the radio pulses, $S_\nu \simeq 4$ mJy, together with a central observing frequency of $\nu = 944$ MHz and a distance of $d = 84$ pc, we obtain a lower limit on the brightness temperature of $T_b \gtrsim 1.8 \times 10^{12}~{\rm K}$. It is worth noting that the assumed source size (\(l = 2R_\star\)) is likely much larger than the true emitting region.  For example, coherent radio emission such as ECM emission is expected to be strongly beamed and to originate from a region occupying only a small fraction of the stellar disk or magnetosphere. Since \(T_b \propto l^{-2}\), any reduction in the true source size would substantially increase the inferred brightness temperature. The actual brightness temperature may therefore be significantly higher than the value quoted above. Even under the full-disk assumption, however, the inferred brightness temperature reaches or exceeds the inverse-Compton limit for incoherent synchrotron or gyrosynchrotron emission \citep{1969ApJ...155L..71K}, and therefore requires a coherent emission mechanism.

Among the known coherent emission processes, ECM emission provides the most natural framework for explaining the observed radio phenomenology. The ECM mechanism naturally produces nearly 100\% circular polarisation, strong directional beaming, and pronounced rotational modulation, all of which are consistent with the observed properties of ASKAP J181335-604720. Within the ECM framework, the characteristic emission frequency is directly related to the local magnetic field strength in the emitting region through the electron cyclotron frequency,
\begin{equation}
\nu_{\rm c} \simeq 2.8\,{\rm MHz}\, B({\rm G}),
\label{eq:cyclotron}
\end{equation}
where $B$ is the magnetic field strength in Gauss. Taking the upper end of the observed frequency range, $\nu = 1080$ MHz, fundamental ECM emission implies a magnetic field strength of $B \sim 390$ G, whereas first harmonic emission corresponds to $B \sim 190$ G. In either case, these values should be regarded as lower limits on the magnetic field strength in the radio-emitting region. 

The magnetic field strengths inferred from the ASKAP band are lower than the kilogauss-level surface magnetic fields commonly measured in active M dwarfs \citep[e.g.,][]{2007ApJ...656.1121R,2019A&A...626A..86S}, but this difference is physically expected. The electron cyclotron frequency constrains the local magnetic field strength at the radio-emitting site, whereas magnetic field measurements based on Zeeman broadening or Zeeman--Doppler imaging generally probe the photospheric field or the large-scale surface field. Moreover, ECM emission may originate above the stellar surface in an extended magnetosphere, where the magnetic field has declined from its photospheric value. Thus, the magnetic field strengths implied by the ASKAP frequencies are physically compatible with an M dwarf magnetic environment, even though the present observations do not constrain the high frequency cutoff of the emission or the maximum magnetic field strength of the star.

We also compare ASKAP J181335-604720 with the Güdel--Benz relation, which describes the empirical radio--X-ray correlation of coronally active stars. We adopt the standard form
\({L_X}/{L_{R}} = 10^{15.5\pm0.5}~{\rm Hz}\), appropriate for active dwarfs \citep{1993ApJ...405L..63G,1994A&A...285..621B}. Using the ASKAP Epoch 1 Stokes~\(I\) continuum peak flux density of \(0.70~{\rm mJy~beam^{-1}}\), we estimate a radio luminosity of \(L_{R}=5.91\times10^{15}~{\rm erg~s^{-1}~Hz^{-1}}\). Together with the X-ray luminosity derived above, \(L_X = 2.27\times10^{29}~{\rm erg~s^{-1}}\), this gives \({L_X}/{L_R} = 3.84\times10^{13}~{\rm Hz}\). This is about \(1.9\) dex below the canonical Güdel--Benz value, meaning the source is radio overluminous by a factor of 79 compared to the expected radio luminosity for its X-ray emission (see Figure~\ref{fig:A3}). Even relative to the lower edge of the relation, it remains 25 times overluminous. If the peak radio flux density of the pulses were used instead, the inferred radio luminosity, and hence the deviation from the Güdel--Benz relation, would be even larger. This behaviour is similar to that seen in auroral or ECM candidates \citep{2021ApJ...919L..10P}, and provides an additional argument against ordinary coronal radio emission as the dominant origin of the ASKAP pulses.

The ECM interpretation also provides a self-consistent explanation for the other key observational properties reported here. In the classical flare scenario, radio bursts are generally associated with impulsive magnetic reconnection events, which are often accompanied by optical or ultraviolet flaring. By contrast, ECM emission can be driven by non-thermal electron populations associated with magnetospheric current systems, and can therefore naturally produce high brightness temperature, highly circularly polarised radio pulses without corresponding optical flares \citep[e.g.,][]{1982ApJ...259..844M,2006A&ARv..13..229T,1998JGR...10320159Z}. At the same time, the confinement of the radio emission to narrow and reproducible rotational phase intervals suggests that the emission is organised by a large-scale magnetic geometry that remains sufficiently stable over the observed rotation cycles. However, the differences in pulse amplitude and morphology between the two ASKAP epochs indicate that the emitting region, beaming conditions, or energetic electron population may vary on timescales of days. 

In a simplified picture, ECM emission is generated along auroral magnetic field lines and beamed into a hollow cone oriented approximately perpendicular to the local magnetic field, as in planetary auroral radio emission and in the standard interpretation of pulsed radio emission from ultracool dwarfs \citep[e.g.,][]{1998JGR...10320159Z,2007ApJ...663L..25H,2009ApJ...695..310B,2012ApJ...760...59N}. As the star rotates, the observer's line of sight periodically sweeps through this emission cone, producing short-duration radio pulses at preferred rotational phases. Taken together, these results support an interpretation in which the radio emission from ASKAP J181335-604720 arises from an auroral-like ECM process organised by a large-scale stellar magnetic geometry and modulated by stellar rotation.

\section{Implications for coherent radio emission in M dwarfs}

Coherent, highly polarised radio emission is now known to occur across a broad range of low-mass stars, from earlier M dwarfs to ultracool dwarfs \citep[e.g.,][]{2007ApJ...663L..25H,2015Natur.523..568H,2017ApJ...846...75P,2019ApJ...871..214V,2021NatAs...5.1233C}. In ultracool dwarfs, periodic radio pulses with high circular polarisation have long been interpreted as auroral ECM emission associated with large-scale magnetospheric current systems, analogous to planetary auroral radio emission \citep{1998JGR...10320159Z,2000RvGeo..38..295B,2015SSRv..187...99B}. For earlier M dwarfs, however, the physical origin of coherent radio emission can be more ambiguous, because both flare-driven coronal activity and magnetospheric processes may contribute \citep{2021NatAs...5.1233C, 2021ApJ...919L..10P, 2025ApJ...990L..32L}. ASKAP J181335-604720 adds to this broader picture as a well-constrained early--mid M dwarf case in which the radio polarisation, rotational phase behaviour, and strictly simultaneous non-detection of optical flares jointly support a magnetospheric coherent emission interpretation.

This result highlights the diagnostic value of simultaneous radio and optical observations for distinguishing between flare-associated radio bursts and rotationally modulated magnetospheric emission in M dwarfs. Optical rotational modulation alone can reveal surface inhomogeneities or magnetic activity, but it cannot determine the physical origin of the coherent radio emission. Conversely, highly circularly polarised radio pulses recurring at preferred rotational phases strongly indicate a coherent process, but simultaneous optical photometry is needed to test whether the radio emission is accompanied by optical flares.

Future observations will be important for testing the magnetospheric ECM interpretation and for constraining the geometry and stability of the emitting region. Long-term radio monitoring can determine whether the radio pulse phases persist over many rotation cycles, and can quantify changes in pulse amplitude, morphology, polarisation, and duty cycle across different epochs. Broad-band radio observations extending below and above the ASKAP band would be particularly valuable for identifying possible low or high frequency cutoffs. Such measurements would constrain the range of magnetic field strengths sampled by the ECM source region and test whether the ASKAP detection represents only part of a broader coherent radio spectrum.

Complementary optical and infrared spectroscopy would refine the stellar parameters and activity diagnostics of ASKAP J181335-604720. High-resolution spectroscopy and, where feasible, spectropolarimetry would be especially useful for constraining the surface magnetic field and large-scale magnetic topology, allowing a more direct comparison with the magnetic field strength inferred from the radio emission. Continued simultaneous radio and optical monitoring will also be important for testing whether future radio pulses remain unaccompanied by optical flares, and for further distinguishing rotationally modulated magnetospheric emission from flare-associated radio activity.

\section{Summary}

In this work, we present a clear observational case for coherent, auroral-like radio emission from the early--mid M dwarf ASKAP J181335-604720, based on strictly simultaneous ASKAP radio observations and TESS optical photometry. Our main results are as follows:

\begin{enumerate}
    \item ASKAP detects recurrent short-duration radio pulses from ASKAP J181335-604720 across 800--1088 MHz, with peak circular polarisation fractions approaching 100\%. The emission consists of a dominant main pulse and a weaker secondary component.

    \item The simultaneous TESS light curve exhibits a smooth and highly repeatable quasi-sinusoidal modulation with a period of $P = 5.607 \pm 0.003$~h, which we interpret as the stellar rotation period. When folded on this period, the radio pulses are confined to narrow phase intervals and recur at preferred rotational phases. Similar phase behaviour is also seen in an independent ASKAP epoch separated by several days, although the pulse amplitudes vary between epochs.

    \item No contemporaneous optical flares are detected at the epochs of the radio pulses in the strictly simultaneous TESS photometry. Together with the high circular polarisation and rotational phase confinement of the radio emission, this strongly disfavors a purely stochastic flare-driven origin for this source.

    \item Under conservative assumptions, the inferred brightness temperature yields a lower limit of $T_b \gtrsim 1.8 \times 10^{12}$~K, which exceeds the inverse-Compton limit for incoherent emission and therefore requires a coherent radiation mechanism. We interpret the radio emission as ECM radiation arising from a large-scale magnetosphere, with the observed emission frequencies implying magnetic field strengths of at least a few hundred Gauss in the radio-emitting region.
\end{enumerate}

We have also considered alternative coherent emission scenarios, including star-planet interaction driven ECM emission and long-period radio transient phenomena. However, the present data provide no independent evidence for a close-in planetary companion. The TESS light curve shows smooth rotational modulation rather than an obvious planetary transit signature, and the radio recurrence period is consistent with the optical rotational modulation of the M dwarf, with circularly polarised emission recurring across independent ASKAP epochs. The present observations therefore do not require, and do not preferentially support, a star-planet interaction or long-period transient interpretation.

Overall, ASKAP J181335-604720 provides a clean, strictly simultaneous radio--optical case of coherent radio emission from an early--mid M dwarf. The observed properties are most naturally interpreted as rotationally modulated coherent magnetospheric emission rather than flare-associated activity. More broadly, this work highlights the diagnostic power of simultaneous radio and optical observations for distinguishing magnetospheric coherent emission from classical coronal flaring, and illustrates the potential of wide-field radio surveys such as EMU \citep{2021PASA...38...46N} to uncover new examples of coherent radio emission from nearby low-mass stars.

\acknowledgments

This work was supported by the National SKA Program of China (Grant No. 2025SKA0120103) and the National Natural Science Foundation of China (NSFC; Grant No. 12133002). We thank Tara Murphy and members of her group at the University of Sydney for helpful discussions and valuable comments that improved this work. This scientific work uses data obtained from Inyarrimanha Ilgari Bundara, the CSIRO Murchison Radio-astronomy Observatory. We acknowledge the Wajarri Yamaji People as the Traditional Owners and native title holders of the Observatory site. CSIRO’s ASKAP radio telescope is part of the Australia Telescope National Facility and is funded by the Australian Government, with support from the National Collaborative Research Infrastructure Strategy. This paper includes data collected with the Transiting Exoplanet Survey Satellite (TESS), obtained from the Mikulski Archive for Space Telescopes (MAST) at the Space Telescope Science Institute (STScI). Funding for the TESS mission is provided by the NASA Explorer Program. This work has made use of data from the European Space Agency (ESA) mission Gaia, processed by the Gaia Data Processing and Analysis Consortium (DPAC). Funding for the DPAC has been provided by national institutions, in particular those participating in the Gaia Multilateral Agreement.

\appendix
\section{Additional Figures}

\setcounter{figure}{0}
\renewcommand{\thefigure}{A\arabic{figure}}
\renewcommand{\theHfigure}{A\arabic{figure}}

\begin{figure}[ht!]
\center

\includegraphics[width=0.43\textwidth]{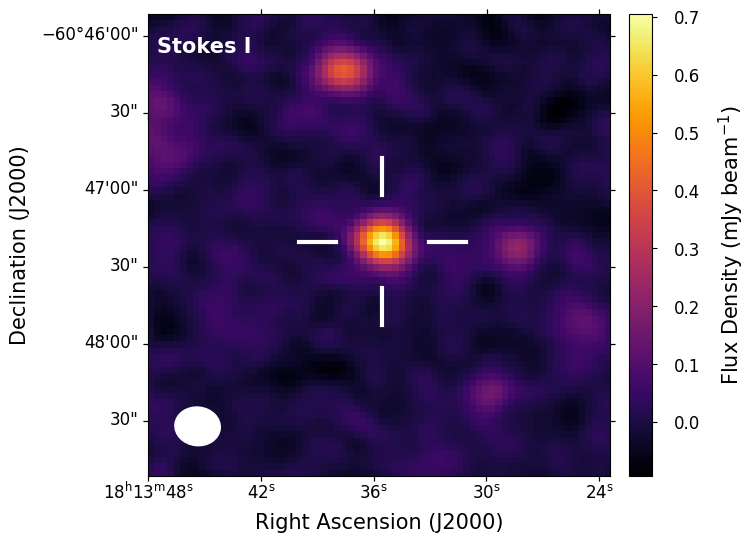}
\includegraphics[width=0.43\textwidth]{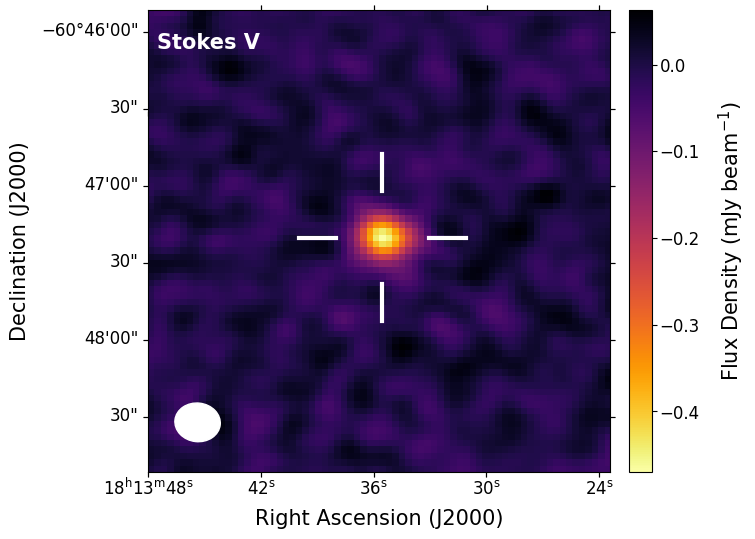}

\center
\caption{Time and frequency averaged ASKAP radio continuum images of ASKAP J181335-604720 in Stokes $I$ and Stokes $V$ (Epoch 1), constructed using the full observing bandwidth of 800--1088 MHz and the entire integration time.}
\label{figa1}
\end{figure}

\begin{figure*}[ht!]
\centering
\includegraphics[width=0.9\textwidth]{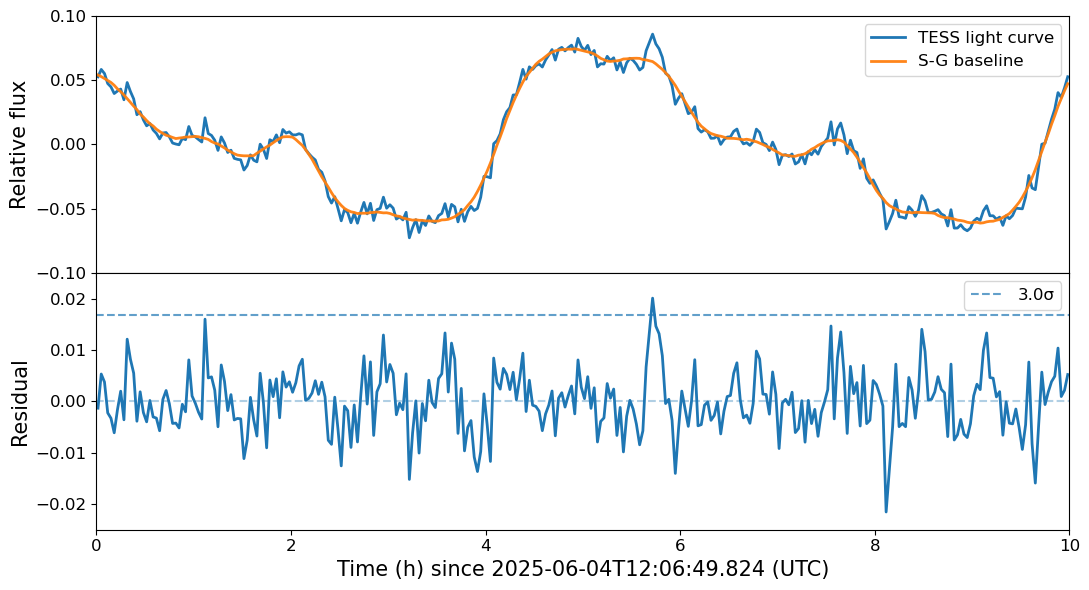}
\caption{Residual light curve after removal of rotational modulation. Top: The strictly simultaneous TESS light curve, shown as relative flux, during the 10 h interval overlapping the ASKAP radio observations. The long-term trend is modelled using a Savitzky--Golay (S--G) filter (orange curve), with a window length of 60 minutes. Bottom: Residual light curve obtained after subtracting the S--G baseline. The dashed horizontal line marks the $3\sigma$ threshold.}
\label{fig:A2}
\end{figure*}

\begin{figure*}[ht!]
\centering
\includegraphics[width=0.6\textwidth]{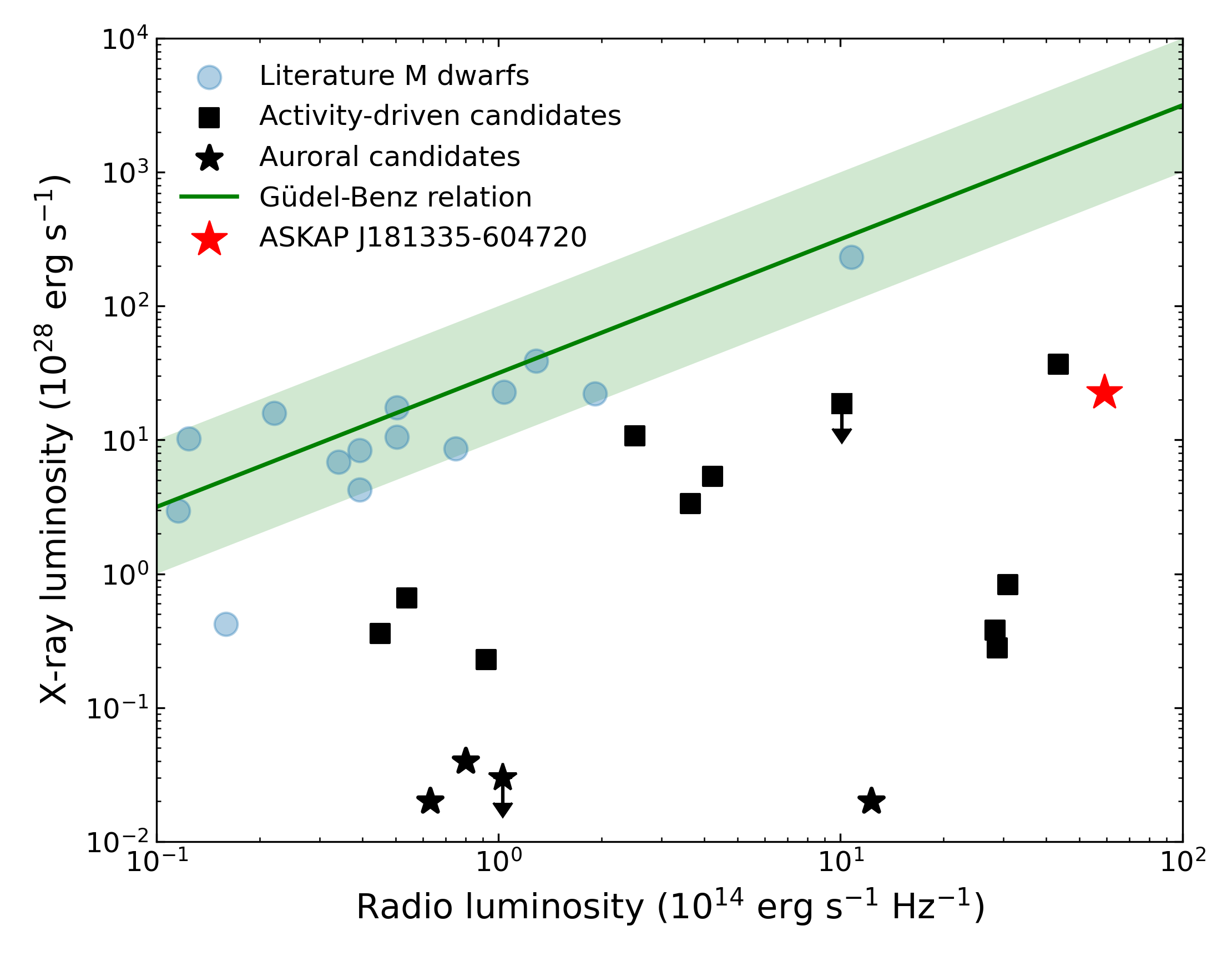}
\caption{Güdel--Benz diagram including ASKAP J181335-604720 and other M dwarfs. 
The red star marks ASKAP J181335-604720, while the blue circles, black squares, and black stars denote literature M dwarfs, activity-driven candidates, and auroral candidates from \citet{2021ApJ...919L..10P}, respectively. The green solid line shows the canonical Güdel--Benz relation, \(L_X/L_{R}=10^{15.5}~{\rm Hz}\), and the shaded region indicates the \(\pm0.5\) dex scatter \citep{1993ApJ...405L..63G,1994A&A...285..621B}.
}
\label{fig:A3}
\end{figure*}

\bibliographystyle{aasjournal}
\bibliography{wangszGC}

\newpage

\end{document}